\documentclass[pre,preprint,amsmath,amssymb,superscriptaddress]{revtex4-1}

\usepackage{wasysym}
\usepackage{graphicx}
\usepackage[TABBOTCAP]{subfigure}
\usepackage{xfrac}

\usepackage{color}
\usepackage{soul}

\begin{document}

\title{Positional ordering of hard adsorbate particles in tubular nanopores}
  \author{P\'eter Gurin}
  \affiliation{Institute of Physics and Mechatronics, University of Pannonia, P.O. Box 158, Veszpr\'em H-8201, Hungary} 
  \author{Szabolcs Varga}
  \affiliation{Institute of Physics and Mechatronics, University of Pannonia, P.O. Box 158, Veszpr\'em H-8201, Hungary}
  \author{Yuri Mart\'{\i}nez-Rat\'on}
  \affiliation{Grupo Interdisciplinar de Sistemas Complejos (GISC), Departamento de Matem\'aticas, Escuela Polit\'ecnica Superior, Universidad Carlos III de Madrid, Avenida de la Universidad 30, E-28911, Legan\'es, Madrid, Spain}
  \author{Enrique Velasco}
  \affiliation{Departamento de F\'{\i}sica Te\'orica de la Materia Condensada, Instituto de F\'{\i}sica de la Materia Condensada (IFIMAC) and Instituto de Ciencia de Materiales Nicol\'as Cabrera,  Universidad Aut\'onoma de Madrid, E-28049 Madrid, Spain}

\date{\today}

\begin{abstract}
  The phase behaviour and structural properties of a monolayer of hard
  particles is examined in such a confinement, where the adsorbed
  particles are constrained to the surface of a narrow hard
  cylindrical pore. The diameter of the pore is chosen such that only
  first and second neighbour interactions occur between the hard
  particles. The transfer operator method of Percus and Zhang [Mol.
  Phys., 69, 347 (1990)] is reformulated to obtain information about
  the structure of the monolayer. We have found that a true phase
  transition is not possible in the examined range of pore diameters.
  The monolayer of hard spheres undergoes a structural change from
  fluid-like order to a zigzag-like solid one with increasing surface
  density. The case of hard cylinders is different in the sense that a
  layering takes place continuously between a low density one-row and
  a high density two-row monolayer. Our results reveal a clear
  discrepancy with classical density functional theories, which do not
  distinguish smectic-like ordering in bulk from that in narrow
  periodic pores.
\end{abstract}

\maketitle

     \section{Introduction}

     There is a growing number of experimental and theoretical studies
     to understand the nature of glassy behaviours, jamming properties
     and phase transitions of confined nanoparticles~\cite{%
       Gelb-Gubbins-Radhakrishnan-SliwinskaBartkowiak_RepProgPhys_2000,%
       Koga_NATURE_2001,%
       Koga-Tanaka_JCP_2006,%
       Lowen_JPhysCondMat_2009,%
       Bowles_PRL_2009,%
       Franosch-Lang-Schilling_PRL_2012,%
       Bowles_PRL_2013,%
       Lang-Franosch-Schilling_JCP_2014,%
       Wensink-Lowen..._EurPhysJ_2013,%
       Leferink...Lekkerkerker_EurPhysJ_2013,%
       Huber_JPhysCondMat_2015,%
       delasHeras_JCP_2015,%
       Robinson-Godfrey-Moore_PhysRevE_2016..2,%
       Wu..._JPC-B_2017%
     }. This is mainly due to the progress made in the experimental
     realization of nanoconfined colloidal systems where the
     confinement reduces the dimensionality of the system in such an
     extent that almost two- and even one-dimensional systems can be
     made.  This can be done with confining the particles between two
     parallel plates ~\cite{Radhakrishnan_PRL_2002} and with
     absorption of the particles into tubular nanopores~\cite{%
       Lohr..._PRE_2010,%
       Bharti..._JAmChemSoc_2012,Jiang_AngewChemIntEd_2013%
     }. The order of phase transitions often changes, the systems
     become frustrated and even new structures appear in the confined
     geometries as a result of competition between particle-particle
     and particle-wall interactions. In the case of slit-like pore
     (parallel plates) it is often found that the first order nature
     of the phase transitions such as the fluid-solid weakens with the
     decreasing pore width and a Kosterlitz-Thouless type continuous
     transition emerges in very narrow pores~\cite{%
       Fortini-Dijkstra_JPhysCondMat_2006,%
       Peng..._PRL_2010
     }.  Similar phenomenon happens in the cylindrical pore, too, with
     a difference that no signs of the bulk first order transitions
     can be observed below a critical radius of the pore because the
     system becomes practically one-dimensional (1d) where the
     particles are not allowed to pass each other~\cite{%
       Gordillo_JCP_2006,%
       Gordillo_PRE_2009,%
       Koga_PNAS_2015%
     }.  Moreover, new types of orientationally and positionally
     ordered structures emerge with the positional restriction such as
     the triatic, tetratic, hexatic and helical arrangements~\cite{%
       Wojciechowski_Frenkel_CompMethSciTech_2004,%
       Donev-Stillinger-Torquato_PRB_2006,%
       Yue_PCCP_2011,%
       Mughal...Hutzler_PRE_2012,%
       Yamchi-Bowles_PRL_2015,%
       Fu-et.al_SoftMatter_2016,%
       Glotzer_PhysRevX._2017%
     }.  In confined liquid crystals the phase behaviour is even
     richer due to the additional orientational freedom. However,
     similar trends can be observed in the phase behaviour of confined
     spherical particles as the phase transitions between different
     mesophases can be suppressed with geometrical
     confinements~\cite{%
       Dijkstra_JCP_2003,%
       Grigoriadis..._ACSNano_2011,%
       Calus_PRE_2012%
     }.

     In the geometrically frustrated systems, Monte Carlo and
     molecular dynamics simulations can be inefficient and very slow
     to find the equilibrium structures due to the slowing down of the
     dynamics and the presence of only few numbers of microstates
     connecting the competing structures. As a result they often
     predict metastable phases and overestimate the order of the phase
     transitions~\cite{%
       Huang..._JCP_2009,%
       Gurin_PRE_2016,%
       Fu-et.al_SoftMatter_2017..1,%
       Khadilkar-Escobedo_SoftMatter_2016%
     }.  In addition to this, the predictions of mean field and
     density functional theories can be even worse as they do not properly
     include the effect of anisotropic and long ranged correlations.
     To remedy the above problems, exactly solvable models can
     be the guideline for both simulation and density functional
     studies. In this regard exact results are available only for
     lattice-models and for some quasi-one-dimensional (q1d) continuum
     models~\cite{%
       Lieb,%
       Abe-Koga_JPhysSocJap_2012,%
       Nath-Dhar-Rajesh_EPL_2016,%
       Mandal-Rajesh_PhysRevE_2017,%
       Wojciechowski_JCP_1982,%
       Wojciechowski_JPhysAMathGen_1983,%
       Forster-Mukamel-Posch_PRE_2004,%
       Kantor_EPL_2009,%
       Khandkar_PRE_2017%
     }. It is also helpful that the true phase transition is forbidden
     in 1d system if the pair potential is short ranged~\cite{%
       vanHove_PHYSYCA_1950,%
       Cuesta_JSP_2004%
     }. In confined hard body models the knowledge of densest or close
     packing structures can be also useful. For example, the densest
     structure of hard spheres in cylindrical tube is very rich
     showing achiral and chiral configurations with changing the
     diameter of the cylinder~\cite{%
       Mughal...Hutzler_PRE_2012,%
       Fu-et.al_SoftMatter_2016%
     }.

     In this study we examine the phase behaviour of hard bodies,
     which are adsorbed on the inner surface of the hard cylindrical
     pore. We resort to the transfer operator method (TOM), which
     proved to be very successful for several q1d systems such as the
     system of hard disks between two lines and that of hard spheres
     in cylindrical pores~\cite{%
       Kofke_JCP_1993,%
       Kamenetskiy-Mon-Percus_JCP_2004,%
       Varga-Ballo-Gurin_JStatMechTheorExp_2011,%
       Gurin-Varga_JCP_2013,%
       Godfrey-Moore_PRE_2014%
     } and the system of freely rotating hard objects constrained to a
     straight line ~\cite{%
       Lebowitz-Percus-Talbot_JStatPhys_1987,%
       Gurin-Varga_PRE_2010,%
       Gurin-Varga_PRE_2011%
     }. It is also applied with less success in narrow periodic box to
     mimic the phase behaviour of the bulk two-dimensional
     systems~\cite{Kofke_JCP_1993}. In these studies the transfer
     operator is constructed in such a way that the positional freedom
     along the channel is integrated out completely and the transverse
     or the orientational freedom of particles kept in the integral
     operator.  This method serves the exact Gibbs free energy and the
     transverse positional or the orientational distributions of the
     particles for systems with only first neighbour interactions. The
     extension of the method has been done recently for wider pores,
     where first and second neighbour interactions are present~\cite{%
       Gurin-Varga_JCP_2015,%
       Godfrey-Moore_PRE_2016%
     }. The fluid of hard squares confined inside a two-dimensional
     infinite channel defined by two hard walls was also studied
     within the TOM and density functional theory \cite{%
       GonzalezPinto-MartinezRaton-Varga-Gurin-Velasco_JPhysCondMat_2016}.
     The results from both theories compared very well. However the
     behaviour of this system is completely different from that of the
     present system, due to the different transverse boundary
     condition used. In the work of Percus and
     Zhang~\cite{Percus-Zhang_MolPhys_1990}, the relative coordinates
     are used to construct the transfer operator of hard squares in
     periodic box with first and second neighbor interactions between
     particles. Their method was used to obtain the equation of state
     of hard squares to see the deviation from the strictly 1d system
     of hard rods and to see the similarity with the two-dimensional
     system of parallel hard squares. They found that there is no
     thermodynamic singularity, i.e. true phase transition cannot
     occur in their system.

     We follow the route of~\cite{Percus-Zhang_MolPhys_1990} with a
     modification of the transfer operator which serves both the exact
     thermodynamics and the distribution function of the first
     neighbors, too. With the new TOM we show that hard spheres form a
     zigzag structure while hard cylinders align into two rows with
     increasing pressure. It is important to emphasize that the
     structural transitions from a low density fluid phase to these
     high density structures are smooth without any thermodynamic
     singularity. Our results are in contrast with some recent density
     functional and Monte Carlo studies related to similar sytems.  In
     Ref.~\cite{%
       Yuri-Enrique_PRE_2013%
     } a nematic--smectic phase transition was found for oriented hard
     rectangles adsorbed on a cylinder, and in Ref.~\cite{%
       Lowen_PCCP_2018%
     } isotropic--nematic and nematic--smectic phase transitions were
     found for freely rotating hard rectangles on a cylindrical
     surface. In Ref.~\cite{Gurin_PRE_2016} the authors find an
     anomalous structural transition of freely rotating hard squares
     confined into a narrow channel which has quite similar behaviour
     as a first order phase transition. Note that such a behavior is
     not observed in the present models. Furthermore, in contrast to
     the freely rotating hard squares, see Ref.~\cite{Gurin_PRE_2017},
     there is no critical behaviour in the infinite pressure limit.
     Finally, in the present study we also shed light on the failure
     of density functional theories, when they are applied to systems
     in narrow periodic pores.

     \section{The cylindrical pore and the adsorbed hard particles}

     We assume that the adsorbent has cylindrical shape, it is
     infinitely long and its inner surface is smooth. Two types of
     hard particles are inserted into the pore; 1) the hard spheres
     with diameter $\sigma$ and 2) the hard cylinders with length $L$
     and diameter $\sigma$. It is also assumed that the cylinders are
     orientated along the direction of the pore as the pore is very
     narrow. Furthermore, the adsorbate particles are always in
     contact with surface of the pore, but they are allowed to move
     freely on the surface of the adsorbent. This means that the
     system has only axial and circular freedom, while the positional
     freedom is switched off in the radial direction. The diameter of
     the pore is $W=D+\sigma$, where $D$ is a diameter of a cylinder
     on which the center of the adsorbate particles can move. We study
     the phase behaviour of hard spheres and that of hard cylinders in
     very narrow pores, where only first and second neighbour
     interactions are possible. In Fig. 1 we show the cartoons of the
     confined cylinders. Here it is also shown that the system of
     confined hard cylinder corresponds to a system of hard rectangles
     moving in a narrow stripe with the periodic boundary condition,
     where the side lengths of the rectangles are $L$ and $\sigma_e =
     D\arcsin(\sigma/D)$ along the axial and circular directions,
     respectively. The latter one, $\sigma_e$ is a contact distance
     between two cylinders along the circular direction. Note that the
     system of confined hard spheres cannot be mapped easily into a
     stripe where the interacting two-dimensional objects are moving,
     because the projection of the contact distance between two hard
     spheres depends on the width of the pore.
\begin{figure}[h!]
  \includegraphics[width=0.98\columnwidth]{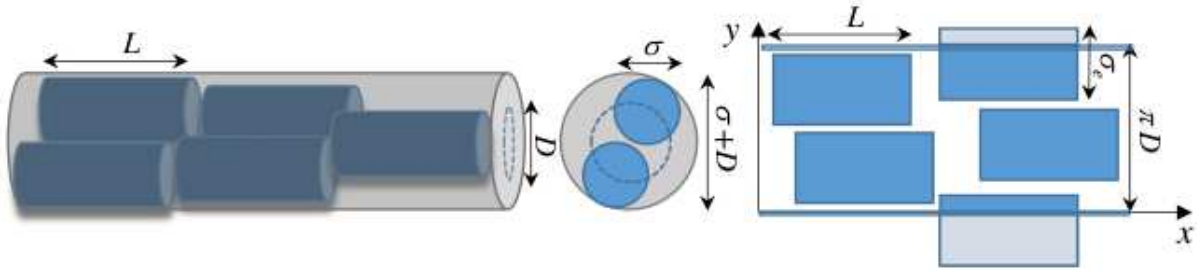}
  \caption{Schematic of the hard cylinders with diameter $\sigma$ and
    length $L$, which are confined to the inner surface of a
    cylindrical pore. The centers of the particles are allowed to move
    on the surface of a cylinder with diameter $D$, which is shown as
    dashed circle.  The unrolled cylindrical surface having side
    lengths $L_x$ and $L_y=\pi D$ along $x$ and $y$ directions,
    respectively, and the projection of hard cylinder particles
    corresponding hard rectangles are shown in the right panel. The
    examined system corresponds to system of hard rectangles with side
    lengths $L$ and $\sigma_e$ , where the particles are moving in a
    periodic narrow box.
  \label{fig:system}}
\end{figure}

     In the following sections we derive the Gibbs free energy, the
     equation of state and the nearest neighbour distribution function
     in the unrolled coordinate frame of the cylindrical surface,
     where the $x$ axis is along the axial direction, while the $y$ one
     corresponds to the circular one. We use the notation $L_x$ and $L_y$
     for the axial and circular dimension of the surface.
 
     \section{The transfer operator method}

     The circular length of our system, $L_y$, is fixed. However, it
     turns out that instead of an isochoric ensemble it is more
     convenient to examine the phase behavior in a mixed
     isobaric-isochoric ensemble, where the independent variables are
     the number of particles inside the pore, $N$, axial force, $F_x$,
     and $L_y$. As the particle-particle and the particle-surface
     interactions are hard repulsive, the temperature ($T$) does not
     play a role in the phase behavior of the system. Consequently,
     the configuration part of the corresponding partition function
     depends on $N$, $ f_x = \beta F_x$ and $L_y$ as follows
\begin{eqnarray}
     Z_N(f_x, L_y)
  =  \frac{1}{N!}
      \int\limits_0^\infty dL_x\; e^{- f_x L_x}
      \int\limits_{0}^{L_y}  \Bigl( \prod_{i=1}^N d \hat y_i \Bigr)
     \int\limits_0^{L_x}     \Bigl( \prod_{i=1}^N d \hat x_i \Bigr) 
     e^{-\beta \sum_{i<j} V(\hat x_i,\hat y_i,\hat x_j,\hat y_j) }
 ,
 \label{eq:Z_def}
\end{eqnarray}
     where $\beta=1/k_B T$, the coordinates of the $i$-th particle are
     $(\hat x_i,\hat y_i)$ and the total interaction energy of the
     system is the sum of the pair potentials, $V$.  Restricting the
     domain of the integration in Eq.(\ref{eq:Z_def}) for a fixed
     order of the particles, $\hat x_1 \leq \dots \leq \hat x_N$, we
     can omit the prefactor $1/N!$ because the interparticle potential
     is symmetric under the permutation of particles. Due to the fact
     that our system is q1d and the interaction is short ranged, i.e.
     the potential energy depends only on the relative coordinates of
     the near neighbours, the integrals of Eq.(\ref{eq:Z_def}) can be
     expressed as a product of transfer operators. To obtain that
     form, as a first step, we change the integration variables to
     relative coordinates as follows: $(\hat x_1,\dots,\hat x_N,L_x)
     \rightarrow (x_0,\dots,x_N)$ and $(\hat y_1,\dots,\hat y_N)
     \rightarrow (y_0,\dots,y_{N-1})$ where $x_0=\hat x_1$, $x_i=\hat
     x_{i+1}-\hat x_i$, $x_N=L_x-\hat x_N$ and $y_0=\hat y_1$,
     $y_i=(\hat y_{i+1}-\hat y_i) \mod L_y$, ($i=1,\dots,N-1$).  With
     these new variables the configurational integral can be written
     as
\begin{align}
     Z_N(f_x, L_y)
  =  \int\limits_{0}^{L_y}  \Bigl( \prod_{i=0}^{N-1} d y_i \Bigr)
     \int\limits_0^{\infty}     \Bigl( \prod_{i=0}^{N} d x_i \Bigr)
     e^{ - \beta \left( \sum_{i=1}^{N-1} V(x_i,y_i) 
                        + \sum_{i=1}^{N-2} V(x_i+x_{i+1},y_i+y_{i+1}) 
                 \right)
         - f_x \sum_{i=0}^{N} x_i }
 .
 \label{eq:Z}
\end{align}
     Here we supposed that only nearest and next nearest neighbour
     interactions are present which gives an upper limit for $L_y$.
     Here and in the followings we mean that the particles are nearest
     neighbours in the distance $x$ only and not in the real distance,
     $r=\sqrt{x^2+y^2}$. Now we rewrite the exponential term of the
     integrand with a special grouping, with left and right boundary
     terms being separated from the bulk one, which is symmetric in
     its variables.  This procedure results in
\begin{align}
    &e^{ - \frac{\beta}{2}V(x_1,y_1) 
         - f_x \left( x_0+\frac{x_1}{2} 
                \right) }
     e^{ - \sum_{i=1}^{N-2}             \left[
           \beta \left(   \frac{1}{2}V(x_i,y_i) 
                        + V(x_i+x_{i+1},y_i+y_{i+1}) 
                        + \frac{1}{2}V(x_{i+1},y_{i+1})
                 \right)
         + f_x \frac{ x_i+x_{i+1}}{2}   \right] }
   \times
 \nonumber 
 \\
    &e^{ - \frac{\beta}{2}V(x_{N-1},y_{N-1} )
                      - f_x  \left( \frac{x_{N-1}}{2} + x_N \right) 
                 }
 .
 \label{eq:V_decomposition}
\end{align}
     Using this expression we can write the partition function as an
     operator product of the middle symmetric terms, while the first
     and the last terms, together with the integrals of their variables,
     act as a linear operator on the operator product as follows
\begin{align}
     Z_N(f_x,L_y)
  =  \varphi(\hat K^{N-2})
 .
 \label{eq:cuesta_form}
\end{align}
     Here $\hat K$ is the transfer integral operator which is defined
     by the kernel
\begin{align}
     K(x,y;x',y')
  =  e^{ -\beta 
         \left( \frac{1}{2}V(x,y) + V(x+x',y+y') + \frac{1}{2}V(x',y') \right)
         - f_x \frac{ x+x'}{2} 
       }
 ,
 \label{eq:K}
\end{align}
     and $\varphi(\hat T)$ is a linear functional,
\begin{align}
     \varphi(\hat T)
  =& \int\limits_{0}^{L_y} dy_0\, dy_1\, dy_{N-1}
     \int\limits_0^{\infty}  dx_0\, dx_1\, dx_{N-1}\, dx_N\,
 \nonumber
 \\
    &e^{ - \frac{\beta}{2}V(x_1,y_1) - f_x \left( x_0+\frac{x_1}{2} \right) 
                 }
     T(x_1,y_1;x_{N-1},y_{N-1})
     e^{ -\frac{\beta}{2}V(x_{N-1},y_{N-1} )
         - f_x  \left( \frac{x_{N-1}}{2} + x_N \right) 
       }
 .
 \label{eq:phi}
\end{align}
     Note that the partition function of several q1d systems can be
     written in the form of Eq.(\ref{eq:cuesta_form}) where the
     exponent of the transfer operator and the linear functional
     $\varphi$ depends on the specific boundary conditions.  However,
     the concrete form of $\varphi$ is not important, because the
     boundary effects do not change the thermodinamic
     properties of the system.

     According to the most general form of the
     Perron--Frobenius--Jentzsch theorem, see \cite{Cuesta_JSP_2004},
     in the thermodynamic limit the Gibbs free energy density is given
     by $\beta g = -\lim_{N\to\infty} \frac{1}{N} \ln Z_N = - \ln
     \lambda_0$, where $\lambda_0$ is the (unique) largest solution of
     the following eigenvalue equation
\begin{equation}
     \lambda_k \psi_k(x,y)
  =  \int_0^\infty \!dx^\prime \int_0^{L_y} \!dy' K(x,y;x',y') \psi_k(x',y')
 .
 \label{eq:eigen_eq_general}
\end{equation}
     In our case the kernel of the transfer operator is defined by
     Eq.~(\ref{eq:K}). The equation of state can be obtained from the
     standard thermodynamic relation between the Gibbs free energy and
     the force, which is now $\rho^{-1} = \partial (\beta g)/\partial
     f_x$, where $\rho= N/L_x$ is the linear density. Moreover, the
     expectation value of any quantity $A(x,y)$, which depends only on
     the relative positions ($x$ and $y$) of nearest neighbour
     particles can be expressed by $\psi_0$ as
\begin{align}
     \langle A \rangle
  =  \int_0^\infty \!dx \int_0^{L_y} \!dy\, A(x,y) |\psi_0(x,y)|^2
 .
 \label{eq:<A>}
\end{align}
     As a special case of Eq~(\ref{eq:<A>}), the distribution function,
     $f(x,y)$, which describes the probability of finding a
     \emph{nearest neighbour} pair in relative position $x$ and $y$ is
     given simply by
\begin{align}
     f(x,y)
     = |\psi_0(x,y)|^2
 .
 \label{eq:f_def}
\end{align}
     Furthermore, we define two nearest neighbour distribution
     functions for $x$ and $y$ distances as
\begin{align}
     f(x)
  =  \int_0^{L_y} \!dy\, f(x,y)
 ,
 \label{eq:fx_def}
\end{align}
     and
\begin{align}
     f(y)
  =  \int_0^\infty \!dx\, f(x,y)
 .
 \label{eq:fy_def}
\end{align}
     The spatial correlation function of a quantity $A$ can be also
     expressed by the eigenfunctions and eigenvalues,
\begin{align}
       G_A(n)
  &:=  \langle A(x_0,y_0) A(x_n,y_n) \rangle - \langle A(x_0,y_0) \rangle \langle A(x_n,y_n) \rangle
  \nonumber \\
  & =  \sum_{k \geq 1} \left( \frac{\lambda_k}{\lambda_0} \right)^n
       \left| \int_0^\infty \!dx  \int_0^{L_y} \!dy \,
       \psi_0(x,y) A(x,y) \psi_k(x,y) \right|^2
 ,
 \label{eq:G_A}
\end{align}
     where $n$ plays the role of a dimensionless discrete measure 
     of the distance between the $0^{\mathrm{th}}$ and
     $n^{\mathrm{th}}$ nearest neighbour pairs (the real distance
     is $n/\rho$ on average).  The absolute value in
     Eq.~(\ref{eq:G_A}) is necessary, however in Eq.~(\ref{eq:<A>})
     and (\ref{eq:f_def}) it is only formal, because in general case
     $\psi_k$ can be complex function, however $\psi_0$ is always real
     and positive.  
     A dimensionless correlation length, $\xi$, is defined by the
     large distance asymptotic behaviour of the correlation function
     via the formula $G_A(n) \sim e^{-n/\xi}$, therefore it follows
     from Eq.~(\ref{eq:G_A}) that $\xi^{-1} =
     -\ln(\lambda_1/\lambda_0)$, because the leading term of the sum
     in Eq.~(\ref{eq:G_A}) corresponds to $k=1$ and all the others
     become negligible as $n\to\infty$.

     We must mention here that the definition of the transfer operator
     is not unique.  If we use the following rearrangement of the
     integrand of Eq.(\ref{eq:Z}) , instead of
     Eq.(\ref{eq:V_decomposition}),
\begin{align}
     e^{ - f_x x_0 }
     e^{ - \sum_{i=1}^{N-2} 
             \left[ \beta
                      \left( V(x_i,y_i) + V(x_i+x_{i+1},y_i+y_{i+1})\right)
                   + f_x  x_i 
             \right] }
     e^{ -\beta V(x_{N-1},y_{N-1} ) - f_x  \left( x_{N-1} + x_N \right) }
 ,
\end{align}
     we obtain the transfer operator of Percus and Zhang, see Eq.~(3.7)
     in ref.~\cite{Percus-Zhang_MolPhys_1990}, but in this case the
     computation of the distribution functions is more complicated,
     since both the left and the right hand side eigenfunctions should
     be needed.  The benefit of the form of Eq.~(\ref{eq:K}) is that
     the kernel of the transfer operator is symmetric and therefore
     the left and right eigenfunctions are the same. As a results it
     is easy to compute the distribution function of the nearest
     neighbor particles.

     \subsection{Hard spheres on a cylindrical surface: the nearest
       neighbour case}

     As a first example we apply the above described formalism for
     hard spheres confined into a narrow cylindrical pore. The
     diameter of the spheres is denoted by $\sigma$, they are absorbed
     on the inner surface of a pore with diameter $D+\sigma$,
     therefore the centers of the spheres can move on a cylinder with
     diameter $D$. In this geometry, the explicit form of the hard
     body interaction gives
\begin{align}
      e^{-\beta V(x,y)}
  =   \theta\left( x^2 + D^2\sin^2(y/D) - \sigma^2 \right)
 ,
 \label{eq:e^V_spheres}
\end{align}
     where $\theta$ is the Heaviside step function. We note that in
     Eq.~(\ref{eq:K}) the prefactor $1/2$ before the nearest neighbour
     potential energy is unimportant in the case of hard body
     interaction, but it is necessary in other cases. We emphasize
     here a substantial difference between the system of hard spheres
     on a cylindrical surface and the system of hard disks in a flat
     2d channel with periodic boundary conditions.  The $x$ projection
     of the contact distance around its minimum value depends linearly
     on $|y-y_{\mathrm{min}}|$ in the case of disks but it is
     quadratic in the case of spheres. Therefore the high axial force
     limiting behaviours of these systems are different. We discuss
     this point it in the Conclusion.\label{disks.vs.spheres}

     When $D<\frac{\sqrt{3}}{2} \sigma$, only nearest neighbour
     interactions take place and the next nearest neighbour term is
     missing from the exponents of Eq.~(\ref{eq:K}), therefore the
     kernel can be written as a product of $x,y$ and $x',y'$ dependent
     factors,
\begin{align}
     K(x,y;x',y')
  =  \theta\left( x^2 + D^2\sin^2(y/D) - \sigma^2 \right)
     e^{ - f_x \frac{x}{2} } 
     \theta\left( x'^2 + D^2\sin^2(y'/D) - \sigma^2 \right)
     e^{ - f_x \frac{x'}{2} } 
     \, .
 \label{eq:K_spheres}
\end{align}
     In this case the solution of the eigenvalue equation is
     straightforward and we find that the only eigenfunction which
     corresponds to a nonzero eigenvalue has the following form
\begin{align}
      \psi_0(x,y)
  =   A\, \theta\left( x^2 + D^2\sin^2(y/D) - \sigma^2 \right)
      e^{- f_x\frac{x-\sigma}{2}}
\end{align}
     where $A$ is a normalization constant. All the other functions
     which are orthogonal to $\psi_0$ correspond to zero
     eigenvalue. The largest eigenvalue is given by
\begin{align}
	       \lambda_0
	=&  \int_{0}^{D\pi} dy \int_0^\infty dx \,
	\theta\left( x^2 + D^2\sin^2(y/D) - \sigma^2 \right) e^{- f_x x}
 \nonumber
 \\
  =&  \frac{1}{f_x} \int_{0}^{D\pi} dy \,
                    e^{-f_x \sqrt{ \sigma^2 - D^2\sin^2(y/D)}}
\end{align}
     The fact that all the other eigenvalues are zero reflects the q1d
     nature of the system.  From~Eq.(\ref{eq:G_A}) we can see that the
     correlation function of any quantity which depends only on the
     relative positions of the nearest neighbour particles is
     identically zero, showing that there are no any correlation
     between the relative positions, i.e. they are statistically
     independent variables. As a consequence, the two-particle
     distribution function can be written as a convolution of the
     nearest-neighbour distribution function $f(x,y)$, similar to the
     case of strictly 1d system of hard rods~\cite{Salsburg_JCP_1953}.

     \subsection{Parallel cylinders on a cylindrical surface: the
       next-nearest neighbour case}

     Here we solve the eigenvalue equation for cylinders with diameter
     $\sigma$ and length $L$ absorbed at the inner surface of a pore
     which diameter is $D+\sigma$. All the cylinders are parallel
     with the pore. Again, as in case of the spheres, the centers of
     the particles can move on a cylinder with diameter $D$ and $L_y =
     D \pi$. Now, due to the hard body interactions, we can write
\begin{align}
      e^{-\beta V(x,y)}
  =   1 - 
      \theta\left( L - x \right) 
      \theta\left( \sigma -  D\,|\!\sin(y/D)| \right)
 \label{eq:theta_spheres}
\end{align}
     When $D<\sigma$, only nearest neighbour interactions take place,
     and the system forms a simple q1d Tonks gas.
     When $\sigma<D<\frac{2}{\sqrt{3}} \sigma$, next-nearest neighbour
     interactions take place, too, but not third neighbour
     interactions, therefore our formalism can be applied. Using
     Eqs.~(\ref{eq:theta_spheres}) and (\ref{eq:K}), it is worth
     writting down Eq.~(\ref{eq:eigen_eq_general}) into two regions of
     $x$.  For $x>L$ (it follows that $\theta(L-x) = \theta(L-(x+x'))
     = 0$) we have
\begin{equation}
\begin{split}
     \lambda_k \psi_k(x,y)
  =  e^{-f_x\frac{x}{2}}
  &  \left[
       \int_0^L \!dx^\prime e^{-f_x\frac{x'}{2}} \int_0^{D\pi} \!dy' 
       \left[1-\theta\left(\sigma-D\left|\sin\frac{y'}{D}\right|\right)\right]
       \psi_k(x',y')
     \right.
\\
  &  \left.
       +
       \int_L^\infty \!dx^\prime e^{-f_x\frac{x'}{2}} \int_0^{D\pi} \!dy'
       \psi_k(x',y')
     \right]
     \; ,
\end{split}
 \label{eq:eigen_eq_x>sigma}
\end{equation}
     while in case $x \leq L$ (it follows that
     $\theta(L-x)=1$) Eq.~\eqref{eq:eigen_eq_general} is reduced
     to
\begin{align}
     \lambda_k \psi_k(x,y)
  =& \left[1-\theta\left(\sigma-D\left|\sin\frac{y}{D}\right|\right)\right]
      e^{-f_x\frac{x}{2}}
      \times
 \nonumber
\\
  &  \left[
       \int_{L-x}^L \!dx^\prime e^{-f_x\frac{x'}{2}} \int_0^{D\pi} \!dy' 
       \left[1-\theta\left(\sigma-D\left|\sin\frac{y'}{D}\right|\right)\right]
       \psi_k(x',y')
     \right.
 \nonumber
\\
  &  \left.
       +
       \int_L^\infty \!dx^\prime e^{-f_x\frac{x'}{2}} \int_0^{D\pi} \!dy'
       \psi_k(x',y')
     \right]
 \label{eq:eigen_eq_x<sigma}
\end{align}
     The lower bound of the first integral is $L-x$ instead of 0
     because three particles can not be closer than $L$, i.e. $x+x'
     \geq L$, when the channel is narrow, $D<2\sigma/\sqrt{3}$.

     If $D\left|\sin(y/D)\right|>\sigma$, one can see that the r.h.s
     of Eq.~\eqref{eq:eigen_eq_x<sigma} in the $x\to L-0$ limit is the
     same as the r.h.s of Eq.~\eqref{eq:eigen_eq_x>sigma} in the $x\to
     L+0$ limit, i.e. $\psi$ is continuous at $x=\sigma$.  Moreover,
     the integrals in Eq.~\eqref{eq:eigen_eq_x<sigma} do not depend on
     $y$, and in Eq.~\eqref{eq:eigen_eq_x>sigma} do not depend neither
     on $x$ nor $y$. Therefore we conclude that the eigenfunctions
     have to be in the following form,
     \begin{equation}
     \psi_k(x,y)
  =  \left\{
     \begin{array}{ll}
       \left[1-\theta\left(\sigma-D\left|\sin(y/D)\right|\right)\right]
       \varphi_k(x) 
                                          & \mathrm{\ if\ } x\leq L 
     \\ 
       e^{-f_x\frac{x-L}{2}} \varphi_k(L)  
                                          & \mathrm{\ if\ } x>L
     \end{array}
     \right.
 \label{eq:psi}
     \end{equation}
     Substituting these forms into
     Eq.~\eqref{eq:eigen_eq_x<sigma} the $y'$ integrals can be
     performed and we can see that $\varphi_k(x)$ is determined by the
     following eigenvalue equation:
\begin{equation}
     \lambda_k \varphi_k(x)
  =  D\pi e^{-f_x\frac{x}{2}}
     \left(
         \frac{e^{-f_x\frac{L}{2}}}{f_x} \varphi_k(L)
       + \varepsilon \int_{L-x}^L \!dx^\prime
         e^{-f_x\frac{x'}{2}} \varphi_k(x')
     \right)
 \label{eq:eigen_eq_phi}
\end{equation}
     where we used the notation 
\begin{equation}
     \varepsilon
  =  1-\frac{2}{\pi}\arcsin\frac{\sigma}{D}
 \label{eq:eps}
\end{equation}
     Differentiating  Eq.~(\ref{eq:eigen_eq_phi}) with respect to $x$ and
     substituting into itself we obtain the following differential
     equation:
\begin{equation}
     \varphi''_k(x) - (f_x \alpha_k)^2 \, \varphi_k(x)
  =  0
 \label{eq:eigen_diff_eq}
\end{equation}
     where
\begin{equation}
     \alpha_k^2
  =  \frac{1}{4} 
     - \left( \frac{D\pi\varepsilon}{f_x\, \lambda_k} 
              e^{-f_x\frac{L}{2}} \right)^2
 \label{eq:alpha}
\end{equation}
     It follows that the solution has the following form: 
\begin{equation}
     \varphi_k(x)
  =  A_k^+ e^{f_x \alpha_k x} + A_k^- e^{-f_x \alpha_k x}
 .
 \label{eq:solution}
\end{equation}
     The integration constants $A_k^+$ and $A_k^-$ can be
     determined by substituting \eqref{eq:solution} into
     \eqref{eq:eigen_eq_phi} obtaining a system of homogeneous linear
     equations.
\begin{align}
     e^{-f_x \alpha_k L} 
     \left(
       1 - \frac{\varepsilon}{\frac{1}{2}+\alpha_k}
     \right) A_k^-
     +
     e^{f_x \alpha_k L} 
     \left(
       1 - \frac{\varepsilon}{\frac{1}{2}-\alpha_k}
     \right) A_k^+
 &=  0
 \label{eq:hom_lin_eq_for_A.1}
\\
     \lambda_k A_k^-
     -
     D\pi\varepsilon \frac{e^{-f_x\frac{L}{2}}}{f_x} \frac{e^{f_x \alpha_k L}}{\frac{1}{2}-\alpha_k} A_k^+
 &=  0
 \label{eq:hom_lin_eq_for_A.2}
\end{align}
     The requirements of the nontrivial solution is that the
     determinants of the 2 by 2 matrix of the coefficients has to be
     zero. It gives the eigenvalues of the transfer operator through
     the following equation
\begin{equation}
     \tanh( f_x \alpha_k L)
  =  2\alpha_k \frac{ 
                \left( \frac{1}{4}-\alpha_k^2 \right) - \varepsilon^2 }{ \left( \frac{1}{4}-\alpha_k^2 \right)(1 - 4\varepsilon) + \varepsilon^2 }
 \label{eq:Percus.Eq.for.eigenvalues}
\end{equation}
     However, this exactly corresponds to Eq.~(3.25) of
     Ref.~\cite{Percus-Zhang_MolPhys_1990} , but the eigenfunction is
     different. From Eq.~\eqref{eq:alpha} and
     \eqref{eq:hom_lin_eq_for_A.2} we get that
\begin{equation}
     \frac{A_k^+}{A_k^-}
  =  e^{-f_x \alpha_k L} \sqrt{\frac{\frac{1}{2}-\alpha_k}{\frac{1}{2}+\alpha_k}}
 \label{eq:A+/A-}
\end{equation}
     therefore the eigenfunction can be written as
\begin{equation}
     \varphi_k(x)
  =  A_k^-
     \left(  e^{-f_x \alpha_k x} 
           + e^{ f_x \alpha_k (x-L)} 
             \sqrt{\frac{\frac{1}{2}-\alpha_k}{\frac{1}{2}+\alpha_k}} 
     \right)
 \label{eq:solution.2}
\end{equation}
     where $A_k^-$ is only a normalization constant.

     The eigenvalues can be expressed from Eq.~(\ref{eq:alpha}) as
     follows
\begin{equation}
     \lambda_k
  =  D \pi \varepsilon \frac{e^{-f_x \frac{L}{2}}}{f_x}
     \frac{1}{\sqrt{\frac{1}{4}-\alpha_k^2}}
 ,
 \label{eq:lambda}
\end{equation}
     therefore Eq.~(\ref{eq:Percus.Eq.for.eigenvalues}) determines all
     the eigenvalues of the transfer operator, and
     Eq.~(\ref{eq:solution.2}) gives the corresponding eigenfunctions
     in case when $\lambda_k \neq 0$. We can also get the expectation
     values and the correlation functions of any quantity which
     depends only on the relative positions of the nearest neighbour
     particles, see Eq.~(\ref{eq:G_A}). To understand its special
     behaviour, we examine the solutions of
     Eq.~(\ref{eq:Percus.Eq.for.eigenvalues}) in detail.

\begin{figure}[h!]
     \includegraphics{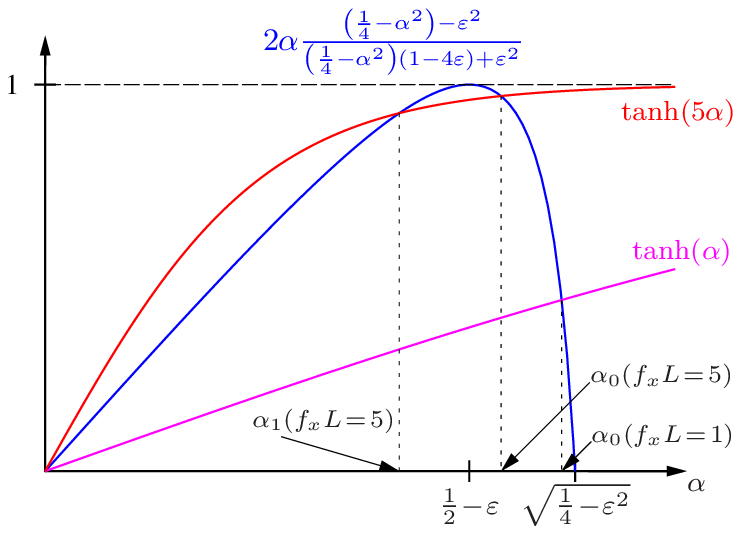}
     \caption{ An examplification of the real solutions of
         Eq.(\ref{eq:Percus.Eq.for.eigenvalues}) for $D^*=1.015$ (i.e.
         $\varepsilon \approx 0.1096$).  We have $\alpha_0 =
         \sqrt{1/4-\varepsilon^2}$ when $f_x = 0$, moreover $\alpha_0$
         is a monotonic decreasing function of $f_x$ and its value
         goes to $1/2-\varepsilon$ as $f_x$ approaches infinity. There
         is only one real solution, $\alpha_0$, when $f_x L <
         \frac{2+4\varepsilon}{1-2\varepsilon}$, and there are two
         different ones, $\alpha_0$ and $\alpha_1$, in other cases.
         The value of $\alpha_1$ also goes to $1/2-\varepsilon$ as
         $f_x$ approaches infinity.
         \label{fig:alpha}}
\end{figure}

     The r.h.s. of Eq.~(\ref{eq:Percus.Eq.for.eigenvalues}), as a
     function of $\alpha$, reaches a local maximum at
     $\alpha=1/2-\varepsilon$, where it is 1, and it is zero at
     $\alpha=\sqrt{1/4-\varepsilon^2}$, see Fig.~\ref{fig:alpha}.
     Furthermore between these points it is monotonically decreasing
     function, therefore Eq.~(\ref{eq:Percus.Eq.for.eigenvalues}) have
     a unique root, $\alpha_0$, in the interval
     $[1/2-\varepsilon,\sqrt{1/4-\varepsilon^2}]$, and $\alpha_0$ is
     monotonic decreasing function of $f_x$ with $\lim_{f_x\to
       0}\alpha_0=\sqrt{1/4-\varepsilon^2}$ and $\lim_{f_x\to
       \infty}\alpha_0=1/2-\varepsilon$. It is easy to see that
     $\alpha_0$ gives the largest eigenvalue, because $\lambda$ is a
     monotonic increasing function of $\alpha^2$ and the r.h.s. of
     Eq.~(\ref{eq:Percus.Eq.for.eigenvalues}) is negative or greater
     than 1 for $\alpha>\sqrt{1/4-\varepsilon^2}$.  Moreover, the
     $\alpha<0$ solutions give the same eigenvalues and eigenvectors
     as the positive ones, thus the proper real solutions are in the
     $[0,\sqrt{1/4-\varepsilon^2}]$ interval. Below $1/2-\varepsilon$
     there is another real solution, $\alpha_1$, when $f_x L \geq
     (2+4\varepsilon)/(1-2\varepsilon)$.  This corresponds to the
     second largest eigenvalue, $\lambda_1$, and $\alpha_1$ has the
     same high axial force limit as $\alpha_0$ has, i.e. $\lim_{f_x\to
       \infty}\alpha_1=1/2-\varepsilon$. All the other solutions of
     Eq.~(\ref{eq:Percus.Eq.for.eigenvalues}) give imaginary
     $\alpha_k$, and they correspond to other smaller eigenvalues,
     which are real because only $\alpha_k^2$ is involved in
     Eq.~(\ref{eq:lambda}). We note that $\alpha_1$ is imaginary, too,
     for $f_x L < (2+4\varepsilon)/(1-2\varepsilon)$, but
     $\alpha_1^2$, and so $\lambda_1$ is analytic function of the
     dimensonless axial force at $f_x
     L=(2+4\varepsilon)/(1-2\varepsilon)$, too.

     The transcendental equation~(\ref{eq:Percus.Eq.for.eigenvalues})
     does not have closed-form solutions, however, for further studies
     of the one row--two row layering transition we can derive an
     approximate solution which can be handled easily. We are
     motivated from two sides. On the one hand getting analytic
     formulas is useful to study the limiting behaviour, on the other
     hand, the numerically precise solution of
     ~(\ref{eq:Percus.Eq.for.eigenvalues}) is difficult, because
     $\alpha_0$ changes its value in a very narrow interval, so when
     $D\to\sigma$, (i.e. $\varepsilon\to 0$) one needs high precision
     for $\alpha_0$. Also in Ref.~\cite{Percus-Zhang_MolPhys_1990} the
     authors give approximate solutions for low and high pressures
     (i.e. axial forces), which are based on the series expansion of
     the $\tanh$ function, thus neither the small nor the large $f_x$
     approximations can describe the most interesting intermediate
     axial force regime when the system changes its behaviour from a
     one-row to two-row layering.  To go beyond their results, here we
     show a more efficient approximation.

     Just the difficulty of the numerical solution of our problem
     gives the key for the analytic approximation. We know that
     $\alpha_0\in[1/2-\varepsilon,\sqrt{1/4-\varepsilon^2}]$,
     therefore $\alpha_0 \approx 1/2$ for arbitrary axial force when
     $D$ approaches to $\sigma$ from above ($\varepsilon\to 0$).
     Therefore we can use the approximation $\tanh(f_x \alpha_0 L)
     \approx \tanh(f_x L/2)$. Moreover, we know that $\lim_{f_x\to
       \infty}\alpha_1 = 1/2-\varepsilon$, therefore, at least in the
     high axial force limit, also the $\tanh(f_x \alpha_1 L) \approx
     \tanh(f_x L/2)$ approximation is valid. As $D$ goes to $\sigma$
     these approximations are better for arbitrary value of $f_x$.
     However, for very small $f_x$ this is not a good approximation.
     But that is clearly the ideal gas limit which is not very
     interesting. What is important, that our approximation is quite
     good for well above $D=\sigma$ when $f_x$ is large and very good
     for low $f_x$, too, in case when $D\gtrapprox \sigma$.  Therefore
     it is not necessary to use series expansion for the $\tanh$
     function, because we get a cubic equation for $\alpha$.  The
     roots of this cubic equation can be used to study the behaviour
     of the system in a very wide $f_x$ range except the ideal gas
     limit.

     Nevertheless, we can do further reliable approximations to get
     analytical formulas which are more manageble than the the roots
     of a cubic equation.  The r.h.s. of
     Eq.~(\ref{eq:Percus.Eq.for.eigenvalues}), and also its
     derivatives change very rapidly with $\alpha$, therefore it can
     not be approximated by a single low order polynomial, but the
     nominator and the denominator are simple second order polynomials
     which change their values slowly, therefore we can use first
     order approximations for them. As we are interested mainly in the
     high $f_x$ behaviour, we use a series expansion around
     $1/2-\varepsilon$ both in the numenator and the denominator. We
     obtain the following quadratic equation in $\alpha$,
\begin{equation}
     \tanh \frac{f_x L}{2}
  =  2\alpha_{0,1} \frac{1-2\alpha_{0,1}}{ (1-2\varepsilon)^2 - (1-4\varepsilon)2\alpha_{0,1} }
\; .
 \label{eq:approx.Eq.for.eigenvalues}
\end{equation}
     The solution of this equation for $\alpha$ can be written as
\begin{equation}
     \alpha_{0,1}
  =  \frac{\varepsilon+(\frac{1}{2}-\varepsilon)e^{f_x L} \pm \sqrt{\frac{1}{4} + (e^{f_x L}-1)\varepsilon(1-2\varepsilon)}}{e^{f_x L}+1}
\; .
 \label{eq:approx.alpha}
\end{equation}
     From this equation we get the limits
\begin{equation}
  \alpha_{0,1}
  \approx \frac{1}{2}-\varepsilon \pm \sqrt{\varepsilon} e^{-f_x \frac{L}{2}}
  \approx \frac{1}{2}-\varepsilon 
 \label{eq:alpha_2layer_p}
\end{equation}
     when $2\varepsilon e^{f_x L} \gg 1$ and
\begin{equation}
     \alpha_{0}
  \approx \frac{1}{2} - \varepsilon^2 e^{f_x L} 
           - \frac{e^{-2 f_x L}}{2}
  \approx \sqrt{\frac{1}{4} - \varepsilon^2 e^{f_x L} }
 \label{eq:alpha_1layer_p}
\end{equation}
     when $2\varepsilon e^{f_x L} \ll 1$. Finally, the last
     approximation in Eq.(\ref{eq:alpha_1layer_p}) is valid when $2\varepsilon^2 \gg e^{-3 f_x L}$.
     Eq.~(\ref{eq:alpha_1layer_p}) is not valid for $\alpha_{1}$
     because it is far from $1/2-\varepsilon$ at low axial force.
     Moreover, it is certainly neither valid for $\alpha_{0}$ when
     $f_x$ is very low, because in our derivation we supposed at the
     beginning that we are far from the ideal gas limit. The case when
     $f_x L \approx -\ln(2\varepsilon)$ defines a special range, where
     neither Eq.~(\ref{eq:alpha_2layer_p}) nor
     Eq.~(\ref{eq:alpha_1layer_p}) can be applied, however
     Eq.~(\ref{eq:approx.alpha}) can be used.

     \section{Results}

     In this section we first present our TOM results for hard spheres
     and show that the close packing zigzag structure develops
     continuously with increasing density (force). After this, the
     packing of hard cylinders is examined, where a layering
     structural change emerges between one-row and two-row fluids.
     The relations of these systems with the well-known
     1d hard rod fluid is also considered as these
     systems become 1d in extreme confining diameters. In
     the case of hard spheres $D=0$ corresponds to the 1d
     limit, while the hard cylinders exhibit this behaviour in the
     range of $0<D< \sigma$. The results of
     Tonks~\cite{Tonks_PhysRev_1936} can be summarized in such narrow
     pores as follows: 1) there is no fluid-solid phase transition as
     the system is one-dimensional with short-range interactions; 2)
     the equation of state can be written down with $f_x =
     \rho/(1-\rho/\rho_{cp})$ (it is called Tonks equations) in the
     whole range of density, where $\rho_{cp}$ is the close packing
     density; and 3) the nearest neighbour probability distribution
     function is decaying exponentially with the distance of the
     neighbors, i.e. $f(x) = \theta(x-l)f(l) e^{-f_x(x-l)}$,
     where $l$ is the length of the hard rod~\cite{Salsburg_JCP_1953}.
     Now we continue with presenting our exact results in wider pores,
     where the following dimensionless quantities are used: 1) $x^* =
     x/\sigma$, $y^* = y/\sigma$, $D^* = D/\sigma$, $f_x^* = f_x
     \sigma$, $\rho^*=\rho\sigma$ for spheres, and 2) $x^* =
     x/L$, $D^* = D/\sigma$, $f_x^* = f_x L$, $\rho^*=\rho L$
     and $\kappa_{_T}^*=\kappa_{_T}/(\beta L)$ for cylinders.

     \subsection{Hard spheres on the surface of cylindrical tube}

\begin{figure}[h!]
     \includegraphics{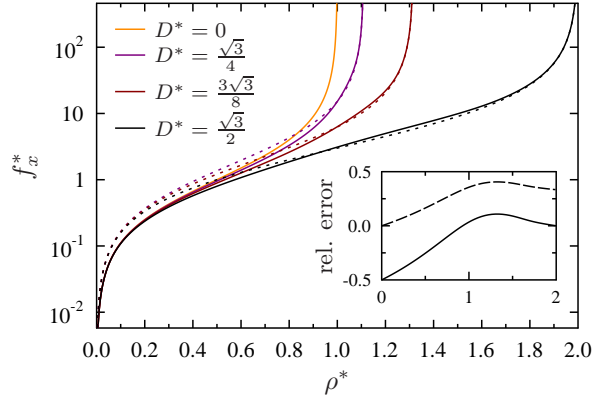}
     \caption{The equation of state (dimensionless axial
       force vs.  1d density) of the spheres confined into a
       cylindrical surface is shown at different pore diameters.
       Continuous curves corresponds to the exact results, while the
       dashed ones is obtained in the high
       axial force limit. The dashed and the continuous
       curves are identical at $D=0$. The deviation of the equation of
       state from the Tonks equation (dashed curve) and from the
       equation of high $f_x$ limit (continuous curve) is
       shown for the widest pore in the inset.
       \label{fig:eos_spheres}}
\end{figure}

\begin{figure}[h!]
     \includegraphics{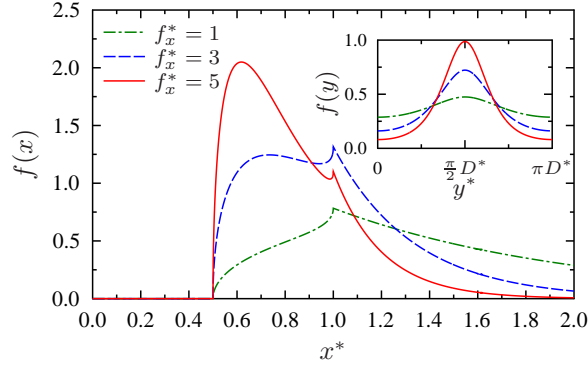}
     \caption{ Nearest neighbor axial and circular distribution
       functions for the spheres confined into a cylindrical surface
       are shown for different axial forces at $D=\sqrt{3}\sigma/2$.
       Axial distribution ($f(x)$) vs. nearest neighbor axial distance
       ($x$) in the main figure, while circular distribution ($f(y)$)
       vs. nearest neighbor circular distance ($y$) in the inset.
     \label{fig:f(x)_f(y)_spheres}}
\end{figure}

     Our results are presented for $0<D^*<\sqrt{3}/2$, where only
     first neighbour interactions are present and zigzag structure
     develops with the increasing density. Note that our formalism
     using the kernel from~Eq.(\ref{eq:K})
     with~Eq.(\ref{eq:e^V_spheres}) remains also valid in wider pores
     of $\sqrt{3}/2<D^*<1$. However, we leave this range for future
     studies, because the situation can be much more complicated as
     chiral structures compete with zigzag structures in the vicinity
     of close
     packing~\cite{Mughal...Hutzler_PRE_2012,Fu-et.al_SoftMatter_2016}
     and analytic solutions cannot be derived from TOM. In
     Fig.~\ref{fig:eos_spheres}
     we show the equation of state of hard spheres at different pore
     diameters from zero to close packing densities. There are two
     reasons why the shape of the curve is changing with $D$: 1) the
     close packing density increases with $D$ as follows: $\rho_{cp} =
     (\sigma^2 - D^2)^{-1/2}$, and 2) it can be shown analytically
     that $f_x = \sfrac{3}{2}\,\rho/(1-\rho/\rho_{cp})$ for $D \neq 0$
     in high axial force limit instead of Tonks equation, which is valid
     only for $D=0$. We show the relative error
     ($1-f_x^{\mathrm{approx.}}/f_x$) of the above equation and the
     Tonks equation as a function of density in the inset of
     Fig.~\ref{fig:eos_spheres} for $D^*=\sqrt{3}/2$.  One can see
     that the Tonks equation is better at low densities, while the
     high axial force limiting equation becomes quite accurate for
     $\rho^*>1$. At intermediate densities ($0.5<\rho^*<1$) both
     equations are inaccurate with about 20--30\% error. This shows
     that the equation of state cannot be written down with a simple
     Tonks-type equation in the whole range of density. The simple
     reason for this is that the system of hard spheres undergoes a
     change from a fluid-like to a zigzag-like structure.  The
     \sfrac{3}{2} prefactor of the equation of state is the
     consequence of zigzag arrangement because $x$ and $y$ coordinates
     couple in such structures. The axial and circular nearest
     neighbor probability distribution function, which are obtained
     from~Eqs.(\ref{eq:fx_def},\ref{eq:fy_def}), justifies the gradual
     structural change with increasing axial force (density) (see
     Fig.~\ref{fig:f(x)_f(y)_spheres}).
     At intermediate densities (see the curve for $f_x^*=1$) one can see
     that $f(x)$ has only one maximum at $x^*=1$ and decays
     exponentially as it happens in the case of Tonks gas. However,
     the fact that $f(x)>0$ in the interval of $1/2<x^*<1$ shows that
     there are some neighbors with $x^*<1$ distance and these
     particles must form zigzag dimer. The structure of $f(y)$ also
     confirms this, as it has a peak when the neighbors are at the
     opposite side of the pore ($y=D \pi/2$).  At higher densities the
     structure of the system substantially deviates from that of 1d
     rods as $f(x)$ becomes two-peaked (see the curve for $f_x^*=3$):
     one peak at $x^*=1$ is the remainder of the 1d hard rod
     structure, while the second one at $x^*<1$ is due to the
     emergence of zigzag clustering. At this density the structure of
     the system is mixture of fluid and zigzag clusters.  At higher
     densities (see the curve for $f_x^*=5$) the majority of the
     particles get into the zigzag cluster, while the fluid-like
     structure is suppressed as $f(x^*=1)$ decreases. At the close
     packing limit ($f_x \to \infty$), only one peak survives in
     $f(x)$ at $x^*=\sfrac{1}{2}$, while $f(x^*=1$) vanishes. It is
     also interesting to consider the fraction of neighbors inside the
     border of $x^*=1$, which is defined as $X_{x<\sigma} =
     \int_0^\sigma f(x) dx$ (see Fig.~\ref{fig:X_{x<sigma}_spheres}).
\begin{figure}[h!]
     \includegraphics{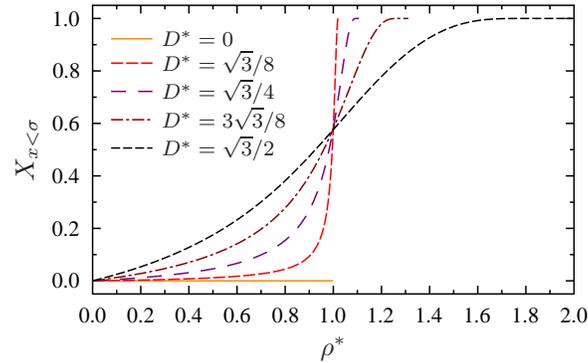}
     \caption{Fraction of nearest neighbors within the axial distance
       $x<\sigma$ vs. 1d density for different pore
       diameters. 
     \label{fig:X_{x<sigma}_spheres}}
\end{figure}
     This serves information about the extent of zigzag structure as
     it is zero in the perfect fluid (1d Tonks gas), while it is one
     in the perfect zigzag structure. Apart from the trivial $D=0$
     limit (corresponding to the system of 1d hard rods),
     $X_{x<\sigma} \to 1$ at the high density limit for any
     $D$. This proves that the TOM accounts for the close packing
     structure of hard spheres at the infinite axial force. It is an
     interesting feature of confined hard spheres that $\rho=1$ is a
     dividing line for all values of $D$, because this is the maximum
     value of density at which the formation of a straight chain of
     hard spheres is still possible.
     Fig.~\ref{fig:X_{x<sigma}_spheres} also shows that the formation
     of zigzag clusters starts at lower densities, while the
     saturation occurs at higher densities with increasing pore diameter.
     Finally we note that the integral equation theory predicts
     similar trends in wider pores, as the positional ordering of hard
     spheres is anisotropic on the outer surface of cylindrical pores,
     too~\cite{Takafumi_JCP_2005}.

     \subsection{Hard cylinders on the surface of cylindrical tube}

\begin{figure}[h!]
     \includegraphics{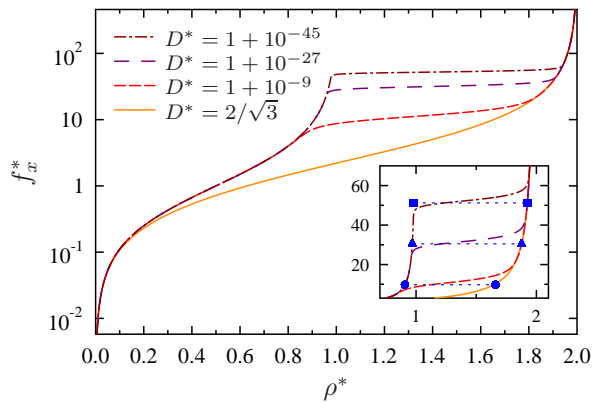}
     \caption{ Equation of state of oriented hard cylinders in a
       cylindrical pore at different pore diameters. The main figure
       is made in log-lin scale, while the inset in lin-lin scale.
       The horisontal blue lines correspond to the one-row to two-row
       transition force, $f_x^* = -\ln 2\varepsilon$.  Symbols at
       the end of these lines denote the densities coming from the low
       end high axial force approximations according to
       Eq.~(\ref{eq:alpha_2layer_p}) and
       Eq.~(\ref{eq:alpha_1layer_p}).
    \label{fig:eos_cylinders}}
\end{figure}

     The examined range of pore diameters can be divided into two
     regions: 1) if $0<D^*<1$, only a row of hard cylinders can form
     and 2) if $1<D^*<\sqrt{2}/3$, two rows of cylinders are allowed
     to form because two hard cylinders can be located at the same
     axial position, $x$. The first case corresponds to the well-known
     1d hard rod fluid even if the particles are allowed to move on
     the circle with radius D, because the axial contact distance is
     always $L$. Therefore one may expect that the 1d hard rod nature
     survives in a pore with $D^*=1+10^{-\epsilon}$, where
     $\epsilon\to\infty$, while two rows develop continuously in the
     possible widest pore ($D^*=2/\sqrt{3}$). To see the
     difference between the two limiting cases, where maximum two rows
     are allowed to form, we present the equation of state at some
     values of $D^*$ in Fig.~\ref{fig:eos_cylinders}.
     One can see that all curves converge to the close packing
     density of two-row structure, $\rho_{cp}^{(2)}$, which is given by
     $\rho_{cp}^{(2)}=2/L$. In addition to this, from
     Eq.(\ref{eq:alpha_2layer_p}) it immediately follows that the
     equation of state is given by $f_x = \rho /
     (1-\rho/\rho_{cp}^{(2)})$ in the high axial force limit, which is
     independent from the pore diameter.  These results indicate that
     the high density structure consists of two rows, where the Tonks
     equation of 1d hard rods is also valid with the appropriate close
     packing density $\rho_{cp}^{(2)}$. The reason why the Tonks
     equation is valid is that there is no coupling between $x$ and $y$
     coordinates of the hard cylinders, which is not the case in the
     zigzag structure of hard spheres.  One can also see that the hard
     cylinders behave very differently at intermediate densities for
     pore diameters very close to the lower limit of $D^*=1$. This
     manifest in a very steep equation of state at the vicinity of the
     axial force $f_x^* = -\ln 2\varepsilon$, where the
     density suddenly changes almost from $\rho_{cp}^{(1)}$ to
     $\rho_{cp}^{(2)}$. Here $\rho_{cp}^{(1)}=1/L$ denotes the close
     packing density of one row fluid. Below this density, as follows
     from Eq.(\ref{eq:alpha_1layer_p}), the equation of states can be
     well approximated by the Tonks equation $f_x = \rho /
     (1-\rho/\rho_{cp}^{(1)})$. From these results it is clear that
     one row of hard cylinders transforms into two rows of them in a
     narrow window of axial force as can be seen in
     Fig.~\ref{fig:eos_cylinders}.

\begin{figure}[h!]
\subfigure[]
{   
    \includegraphics{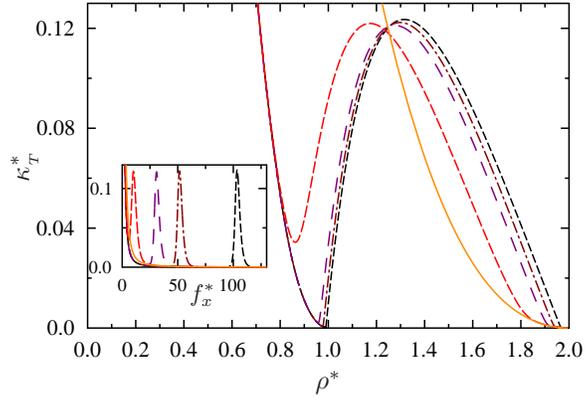}
    \label{fig:kappa_cylinders}
}
\subfigure[]
{   
    \includegraphics{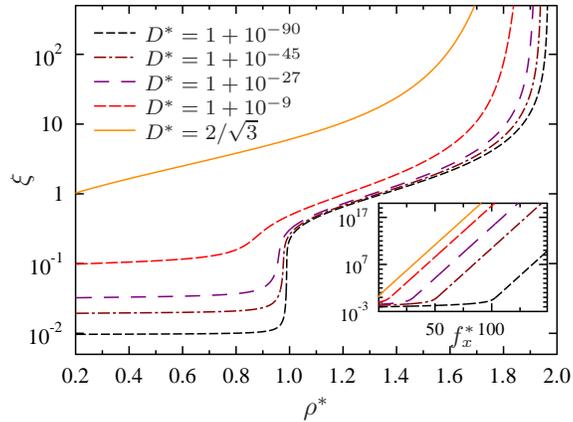}
    \label{fig:xi_cylinders}
}
\caption{(a) Dimensionless isothermal compressibility as a function of
  density for several pore diameters.  (b) Correlation length as a
  function of density.  The insets show the same as the main figures
  as a function of axial force. The color code of the curves are shown
  in (b).\label{fig:kappa-xi_cylinders}}
\end{figure}

     The border between the validity of the high and low force
     approximations, Eqs.(\ref{eq:alpha_2layer_p}) and
     (\ref{eq:alpha_1layer_p}), respectively, is given by $f_x^* =
     -\ln(2\varepsilon)$. The layering cross-over between one-row and
     two-row structures happens at this point. We mention here, that
     the behaviour of the system at this axial force can be
     approximated as a first order transition if the connecting terms
     in the kernel of the transfer operator is omitted. In this
     approximation the transfer operator is not a positive operator,
     therefore the phase transition cannot be excluded
     anymore~\cite{Cuesta_JSP_2004}.  The Gibbs free energy of one row
     fluid becomes identical with that of 1d rods, while that of
     two-row fluid is approximated using the high axial force limit of
     $\alpha$, see Eq.(\ref{eq:alpha_2layer_p}).  Equating the Gibbs
     freeenergies ensures the equality of chemical potentials and
     provides the transition pressures, which gives exactly the border
     between the above low and high $f_x$ approximations, $f_x^* =
     -\ln(2\varepsilon)$.  At this axial force it is easy to determine
     the coexisting densities of the fluids from the corresponding
     equation of states. We show these pairs of densities in the inset
     of Fig.~\ref{fig:eos_cylinders} with blue markers. This inset
     shows the equation of state in linear scale.  It can be seen that
     the transition is shifted upwards according to $f_x^* = -\ln
     2\varepsilon$ as $D\to 2$.  However, there is no genuine phase
     transition in the system, the exact analytic curves show that the
     equation of state is very flat in this region. Therefore it is
     interesting to examine the behaviour of the isothermal
     compressibility, $\kappa_{_T} := -\frac{1}{L_x}\frac{\partial
       L_x}{\partial f_x}= -\rho \frac{\partial^2(\beta g)}{\partial
       f_x^2}$ and the correlaltion length, $\xi^{-1} =
     -\ln(\lambda_1/\lambda_0)$ which can be determined using
     Eq.~(\ref{eq:approx.alpha}). The curves of the resulting lengthy
     formulas are shown in Fig.~\ref{fig:kappa-xi_cylinders}.  The
     compressibility curve, see Fig.~\ref{fig:kappa_cylinders}, has a
     bump at the transition but the height of this bump is constant.
     Here we have no divergence, which is in contrast with some
     other q1d systems, where real critical divergences occur as some
     geometrical parameters approach to a special critical value and
     the axial force goes to infinity, see e.g.
     Ref.~\cite{Gurin_PRE_2017}.  Interesting to see that in the close
     packing limit of the one-row structure the system is very rigid,
     the compressibility becomes very small. The behaviour of the
     correlation length also supports the aboves.  Below the
     structural change the correlations are very weak and the
     correlation length is almost zero, see
     Fig~\ref{fig:xi_cylinders}. This behaviour is very similar to
     that of spheres because the nearest neighbour intaractions
     dominate the behaviour of the system in the case of one-row
     structure, while the next-nearest neighbour interactions hardly
     play any role. However the correlation length starts to diverge
     exponentionally with the axial force as the two-row structure
     builds up at the structural transiton. We can show using
     Eq.~(\ref{eq:alpha_2layer_p}) that $\xi \approx \sqrt{\varepsilon
       e^{f_x^*}}$ above the transition, but there is no any peak in
     $\xi$ at the transition.  Here we note that in a very recent
     manuscript, Ref.~\cite{Hu-Fu-Charbonneau_arXiv_2018}, Hu et. al.
     propose that a sudden structural change without genuine phase
     transition, like the one row--two row layering transition in our
     case, can be associated with eigenvalue crossing or splitting of
     the transfer operator. The largest eigenvalue is certainly
     analytic, guaranteed by the Perron--Frobenius--Jentzsch theorem,
     but the second and third or any other eigenvalues can cross each
     other, causing a non-smooth or non-monotonic behaviour of the
     correlation length. Regarding this context we mention that this
     is not the case in our model. All the eigenvalues can be given by
     Eq.(\ref{eq:lambda}) where $\alpha_k$ are determined by
     Eq.(\ref{eq:Percus.Eq.for.eigenvalues}). We have seen that
     Eq.(\ref{eq:Percus.Eq.for.eigenvalues}) has at most two real
     solutions, which can not cross each other, and examining all the
     imaginary solutions we find that neither of them can cross each
     other, therefore that is also true for all the eigenvalues.

\begin{figure}[h!]
     \includegraphics{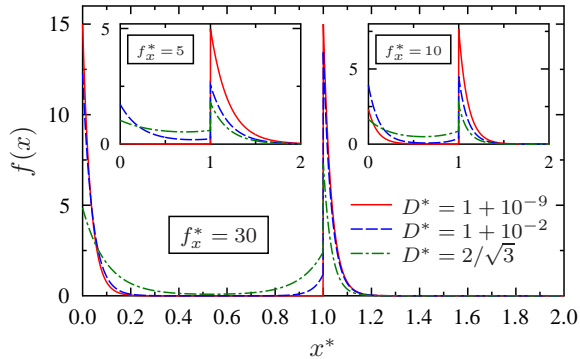}
     \caption{Nearest neighbor axial distribution function as a
        function of axial distance at different axial forces.
     \label{fig:f(x)_cylinders}}
\end{figure}

     The axial nearest neighbour distribution confirms the predictions
     coming from the above analysis, which is shown in wide and
     narrow pores in Fig.~\ref{fig:f(x)_cylinders}.
     In the possible widest pore, $D^*=2/\sqrt{3}$, it can be seen
     that the occupation of the space happens in a way that one-row
     structure does not exist even at very low densities and the close
     packing structure with two rows develops continuously with
     increasing $f_x$. In very narrow pores, $f(x)$ carries the
     feature of 1d hard rods as it is zero for $x<L$ and decays
     exponentially for large distances. Therefore hard cylinders form
     only one row at low and middle densities (see the inset of
     Fig.~\ref{fig:f(x)_cylinders} at $f_x^*=5$).  However, $f(x)$
     becomes peaked at $x=0$ even in the narrowest pores at very high
     axial forces, which proves that the two-row structure develops
     in narrow and wide pores, too. It can be also seen that the peaks
     of $f(x)$ at $x=0$ and $x=L$ becomes sharper and narrower with
     increasing $f_x$ (or density), which is reminiscence of solid
     structures. Further information can be gained from the fraction
     of nearest neighbours ($X$) being in the different regions of the
     axial distance. These regions are the followings: $0<x<L/2$,
     $L/2<x<L$ and $L<x<\infty$. We show $X_{0<x<L/2}$, $X_{L/2<x<L}$
     and $X_{L<x}$ as a function of density for narrow and wide
     pores in Fig.~\ref{fig:X_{...}_cylinders}.
\begin{figure}[h!]
     \includegraphics{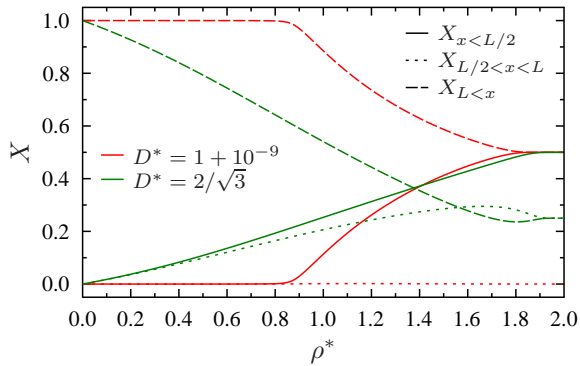}
     \caption{Fraction of nearest neighbors at different regions of
       the axial distance between the neighbors vs. one-dimensional
       density for different pore diameters.
     \label{fig:X_{...}_cylinders}}
\end{figure}
     In narrow pore $X_{L<x}$ is almost one up to about
     $0.9\rho_{cp}^{(1)}$, while the other two are approximately zero.
     This is in agreement with our previous statements that the
     particles form a single row and behaves as a 1d hard rod fluid.
     At the close packing density, one gets that $X_{0<x<L/2}=1/2$,
     $X_{L<x}=1/2$, while $X_{L/2<x<L}=0$. This shows that the first
     right neighbour of a given particle can be either located at the
     same axial position above it or at the distance $x=L$ with equal
     probability as $f(x)$ becomes delta function at the close
     packing.  This suggests that two particles at the same axial
     position form a dimer and the system form a layered structure
     along the axial direction resembling a smectic-like
     configuration.  In wide pore, $D^*=2/\sqrt{3}$, 1d hard rod fluid
     does not exist as $X_{L<x}$ is one only in the ideal gas limit.
     The monotonically increasing feature of $X_{0<x<L/2}$ and
     $X_{L/2<x<L}$ supports the idea that two rows develop
     continuously. As $X_{0<x<L/2}=1/2$, $X_{L/2<x<L}=1/4$ and
     $X_{L<x}=1/4$ at the close packing density, the 50\% of the first
     neighbours can be found at the same axial position as before, but
     only 25\% of the particles is outside of the axial overlap
     region. This indicates that the first neighbours do not form a
     pair and the structure is not layered at the close packing
     density, resembling a columnar-like configuration of particles.
     From these results the close packing structure can not be
     identified with a solid.

     Finally, we touch upon the question of the relation of our
     results with others coming from density functional theories such
     as fundamental measure theory (FMT).  If $D^*<1$ then the
     FMT-based density functionals (DF) recovers the exact Percus
     density functional for hard rods \cite{Yuri-Enrique_PRE_2013}.
     If $D^*>1$, which allows two particles pass each other, the DF is
     not exact. However it can be applied, by imposing appropriate
     periodic boundary conditions, to study configurations of
     particles with density profile varying along the circular axes
     while it is constant along the cylinder axis. As the mean density
     of particles increases the system exhibits a second order
     fluid-columnar phase transition with the transition density not
     depending on $D$ and equal to that of the bulk. Note that at bulk
     the smectic and columar phases of hard squares are identical.
     For certain values of $D$ as the density increases first order
     transitions between columnar phases with different number of
     columns take place with transition densities strongly depending
     on $D$.

     However if we impose the density profile to be constant along the
     circular axes and non-uniform along the axis of cylinder the
     system is completely equivalent to a bulk fluid.  It is clear
     that an extra condition, reflecting the finiteness of the system
     along the transverse direction, should be imposed. The way it can
     be implemented is not clear for us. The unique recipe at the
     level of the grand-canonical one-body density based DF is to
     allow the density to increase up to its maximum close-packing
     value, calculated by taking into account that only two particles
     can fit inside the pore. But if we implement this recipe we
     obtain the same smectic free-energy branch as that of the bulk
     fluid.  Thus the second order fluid-smectic transition only takes
     place if its bulk value fall below the maximum allowed
     close-packing density and in this case the free-energy is always
     bellow that corresponding to the columnar except at the
     bifurcation point.

     Finally if we leave the density profile to vary along both
     transversal and axial directions we find a crystalline phase as a
     local minimum of the DF.  Its relative stability with respect to
     the columnar phase strongly depends on packing fraction and $D$.
     The crystalline phase on the cylinder can even be more stable
     than the smectic only for narrow packing fraction intervals
     usualy located at high densities where a good commensurarion
     between the bulk crystal lattice parameter and the permiter of
     cylinder exists.
     
     In any case if the transition to the smectic finally occurs the
     density profile will be independent on the cylinder radius which
     constitutes a strong drawback of the grand-canonical one-body
     density based DF formalism. Note that if a two-body density based
     DF would be available then the two-body density profile would
     reflects the effect of the system finiteness along the perimeter
     has on transverse particle correlations despite the unchanged
     value of the density profile along the perimeter. On lattice
     systems this kind of DF has been worked out~\cite{%
       Velasco-Tarazona_PRA_1990,%
       Schlijper_JStatPhys_1990,%
       Schlijper_JCP_1991%
     }, so it is mandatory its extension to the continuum. A promising
     route to take into account the finiteness of the system along one
     spatial direction at the level of one-body density based DF could
     consist on the extension of the formalism developed in
     Ref.~\cite{delasHeras_PRL_2014} to obtain a canonical version of
     DF for finite systems from the grand canonical one.

     \section{Conclusions}

     We have reformulated the transfer operator method of Percus and
     Zhang~\cite{Percus-Zhang_MolPhys_1990} to obtain some additional
     information about the thermodynamic properties and the structure
     of quasi-one-dimensional hard body fluids. Our method is based on
     the special grouping of pair interactions to obtain the
     isochoric--isobaric partition function as operator products of
     the transfer operator. In the thermodynamic limit ($N\to\infty$)
     the largest eigenvalue of the transfer operator is the most
     dominant and provides the Gibbs free energy, while the
     corresponding eigenfunction gives information about the local
     structure, because we use a symmetric form of the transfer
     operator. To give an explicit analytic solution, we devised an
     efficient approximation, thus we were able to handle the case
     when the pore is just as wide to allow the accomodation of two
     particles at the same axial position. Our method can be applied
     only for hard particles in cylindrical confinements and stripes
     with periodic boundary conditions where only first and second
     neighbour interactions are present.

     Using our method, we have examined the ordering behaviour of the
     monolayer of hard bodies, which are adsorbed on the surface of
     the hard cylindrical pore. In the case of hard spheres, which are
     adsorbed on the inner surface of the channels, we could perform
     analytic calculations up to pore diameter
     $W=(1+\sqrt{3}/2)\sigma$, which is the upper limit of the first
     neighbor interactions. It is found that the phase behavior of
     confined hard spheres deviates substantially from the strictly 1d
     case, which corresponds to $W=\sigma$. This manifests in the
     formation of zigzag-like structure at high density which can be
     seen in the distribution of first neighbors and the equation of
     state in the vicinity of the close packing density, $\rho_{cp}$.
     This latter one obeys
\begin{equation}
     f_x 
  =  \frac{3}{2} \;\frac{\rho}{1-\rho/\rho_{cp}}
 \; ,
 \label{eq:high_pressure_eos}
\end{equation}
     which is different in a factor of one and half from the well-known
     Tonks's equation of the 1d hard rods, $f_x = \rho /
     (1-\rho/\rho_{cp})$.  The emergence of the extra one and half
     factor is due to the coupling of axial and circular freedoms in
     the zigzag structure. Nevertheless,
       Eq.~(\ref{eq:high_pressure_eos}) differs also from the high
       $f_x$ limiting equation of state of hard disks in narrow stripe
       derived by Godfrey and Moore in
       Ref.~\cite{Godfrey-Moore_PRE_2014}. In their case there is a
       prefactor two instead of one and half like in the case of
       periodic boundary condition. As shown in
       Sec.~\ref{disks.vs.spheres} this is related to the smooth
       behaviour of the $x$ projection of the contact distance in the
       vicinity of its minimum value. This shows that the confined
       hard spheres cannot be mapped onto a system of hard disks in a
       periodic narrow stripe Regarding the hard cylinders, which are
     also adsorbed on the inner surface of the tubular pore, we have
     found that the Tonks's equation of state is exact for
     $\sigma<W<2\sigma$, because the particles are not allowed to pass
     each other in the tube and only first neighbor interactions are
     present. In wider pores $2\sigma<W<(1+2/\sqrt{3})\sigma$, a
     structural change from one-row fluid to two-row fluid
     is observed as the particles are allowed to pass
     each other. This layering change becomes sharper as the pore
     diameter is decreased to the lower limit ($2\sigma$), because
     there is less and less room to form the second row.
     Interestingly, the equation of state of one-row fluid and that of
     two-row one can be described with the Tonks's equation, where the
     corresponding one-row and two-row close packing densities has to
     be used.  Here, the factor of one and half is missing because
     there is no coupling between axial and circular freedoms. We have
     shown that the layering can be seen as a virtual 1st order
     transition using the Gibbs free energies of the purely one-row
     and two-row structures. The lack of true phase transition is due
     to the connecting terms of the transfer operator. Therefore it is
     possible that even simulation studies find this layering change
     to be of 1st order for pore diameter $W=2\sigma+\epsilon$, where
     $\epsilon\to 0$. However, simulation studies can not be
     implemented for such pore diameters. We have found that the local
     structure resembles the solid phase as $W \to
     (1+2/\sqrt{3})\sigma$, while it is smectic-like for $W = 2\sigma
     + \epsilon$, where $\epsilon\to 0$.

     There are two important reasons to use exact methods for systems
     with dimensional restrictions between one and two dimensions.
     First, present-day classical density functional theories (DFT)
     are wrong for quasi-one-dimensional systems, because they predict
     that the phase behavior of cylindrically confined hard cylinders
     is identical to that of the two-dimensional bulk system for
     smectic-like configurations. In addition, DFT predicts phase
     transitions from the fluid to the smectic or columnar phases,
     between columnar phases with different number of columns, and
     finally to the crystalline phase. The relative stability between
     different phases depends on density and pore width. Some of these
     phase transitions occur for
     $2\sigma<W<(1+2/\sqrt{3})\sigma$~\cite{%
       Yuri-Enrique_PRE_2013,%
       Lowen_PCCP_2018%
     }.  The failure of DFT can also be present for other particle
     shapes if particles are constrained on the surface of a cylinder
     or they move on a periodic narrow surface. These unphysical
     results are ultimately due to the unrealistic representation of
     anisotropic correlations in the absence of symmetry-breaking
     external fields~\cite{%
       Schlijper_JStatPhys_1990,%
       Schlijper_JCP_1991%
     }.  Second, even molecular dynamics and Monte Carlo simulations
     can predict 1st order phase transitions in q1d systems such as
     hard spheres embedded into a cylindrical pore, where the
     possibility of a phase transition can be excluded with complete
     certainty~\cite{Cuesta_JSP_2004}. This failure of the simulations
     can be attributed to the poor sampling of the configuration space
     and to the presence of vast amounts of jammed and glassy states.
     To check the reliability of DFT and simulation studies it would
     be useful to extend the transfer operator method for wider pores.

     \acknowledgments

     P.G. and S.V.  would like to acknowledge the financial support of
     the National Research, Development, and Innovation Office (NKFIH
     K124353). Financial support under grants FIS2015-66523-P and
     FIS2013-47350-C5-1-R from Ministerio de Econom\'{\i}a, Industria
     y Competitividad (MINECO) of Spain is acknowledged.

%
\end{document}